\documentclass[conference]{IEEEtran}
\usepackage{cite}
\usepackage{amsmath, amssymb,amsfonts}
\usepackage{mathtools}
\usepackage{mathabx}

\usepackage{graphicx}

\usepackage{dblfloatfix}
\newcommand{\norm}[1]{\left\lVert#1\right\rVert}
\usepackage{afterpage}
\DeclareMathOperator*{\argmin}{arg\,min}
\begin{document}
\title{A Comprehensive Self-interference Model for Single-antenna Full-duplex Communication Systems}

\author{\IEEEauthorblockA{Md Atiqul Islam~and~Besma Smida}
\IEEEauthorblockA{\textit{Department of Electrical and Computer Engineering} \\
\textit{University of Illinois at Chicago}\\
Chicago, USA \\
mislam23@uic.edu,~smida@uic.edu}
\\[-4.5ex]
}

\maketitle
\begin{abstract}
Single-antenna full-duplex communication technology has the potential to substantially increase spectral efficiency. However, limited propagation domain cancellation of single-antenna system results in a higher impact of receiver chain nonlinearities on the residual self-interference (SI) signal. In this paper, we offer a comprehensive SI model for single-antenna full-duplex systems based on direct-conversion transceiver structure considering nonlinearities of all the transceiver radio frequency (RF) components, in-phase/quadrature (IQ) imbalances, phase noise effect, and receiver noise figure. To validate our model, we also propose a more appropriate digital SI cancellation approach considering receiver chain RF and baseband nonlinearities. The proposed technique employs orthogonalization of the design matrix using QR decomposition to alleviate the estimation and cancellation error. Finally, through circuit-level waveform simulation, the performance of the digital cancellation strategy is investigated, which achieves 20 dB more cancellation compared to existing methods.  
\end{abstract}
\section{Introduction}
Full-duplex technology is one of the emerging techniques to achieve higher spectral efficiency, where devices simultaneously transmit and receive at a single frequency. This novel paradigm in improving spectrum usage can increase the throughput by a factor of two compared to the traditional bi-directional half-duplex, namely time-division duplexing and frequency-division duplexing. However, the implementation of full-duplex systems is constrained by the cancellation of self-interference (SI), which is stemming from the transmitter to the receiver hindering the detection of the received signal.
\par 

Recent works \cite{bharadia2013full,choi2010achieving,korpi2017compact,knox2012single,debaillie2014analog,jain2011practical,khaledian2018inherent,alexandropoulos2017joint,ahmed2015all,korpi2014widely,quan2017impacts,balatsoukas2015baseband,korpi2015adaptive,khaledian2018robust,sabharwal2014band,smida2017reflectfx,korpi2014full,anttila2014modeling,sahai2013impact,kiayani2018adaptive,korpi2014feasibility,khaledian2018inherent2,soury2017optimal,khaledian2016power,duarte2012experiment,korpi2018modeling,radunovic2010rethinking,duarte2010full,everett2014passive,syrjala2014analysis,masmoudi2017channel,balatsoukas2018non,iimori2018full,gowda2018jointnull,li2018augmented,sim2018self} have investigated several different system architectures and cancellation techniques to alleviate the SI in the full-duplex receiver. The SI cancellation techniques can be categorized in a tripartite manner: propagation domain isolation, analog/radio frequency (RF) domain suppression and digital cancellation \cite{bharadia2013full}.  The propagation domain isolation or antenna cancellation is accomplished using a combination of antenna directionality \cite{choi2010achieving}, path loss\cite{korpi2017compact}, cross-polarization \cite{knox2012single}, and electrical balance isolator network \cite{debaillie2014analog}. Suppression in analog domain is achieved by subtracting a processed copy of the transmitted signal from the receiver input \cite{jain2011practical,khaledian2018inherent,alexandropoulos2017joint}. However, antenna and analog domain suppression are not sufficient to mitigate the strong SI signal below the weak desired signal\cite{bharadia2013full}. Therefore, the residual SI at the detector input of the receiver chain increases the noise floor, which calls for another cancellation stage in the digital domain.\par 
Digital domain cancellation is accomplished by SI regeneration using digital filters fitted to the detector input through the known transmit data. For digital cancellation techniques,
the effect of transmitter and receiver power amplifier (PA) nonlinearities is analyzed in \cite{bharadia2013full,ahmed2015all}, whereas the impact of in-phase/quadrature (IQ) imbalances and resulting image components is investigated in \cite{korpi2014widely}. Signal models including the phase noise of transmitter and receiver oscillators are presented in \cite{quan2017impacts}. It was observed that the phase noise is a bottleneck for perfect SI cancellation while using two different oscillators for transmitter and receiver. A scenario with common oscillator for both the transceiver sides is also analyzed in \cite{balatsoukas2015baseband}.
\par 
In previous literature, both single-antenna and separate-antenna full-duplex models are investigated. However, using single-antenna as a function of SI cancellation is more attractive because of twofold reasons: first, using two antenna full-duplex system may not achieve any higher throughput than using two antenna in half-duplex MIMO system spatially multiplexing two independent packets at the same time \cite{jain2011practical,bharadia2013full}; second, using single-antenna transceiver system results in a compact design. Existing single-antenna full-duplex antenna and analog cancellation methods typically achieve 50-60 dB cancellation \cite{debaillie2014analog,korpi2015adaptive,khaledian2018robust}, where separate-antenna system provides higher cancellation. For a practical receiver, this limited RF cancellation results in a higher effect of receiver RF and baseband (BB) nonlinearity on the SI signal. \par

\begin{figure*}[!hbtp]
    \centering
    \includegraphics[width=0.9\textwidth]{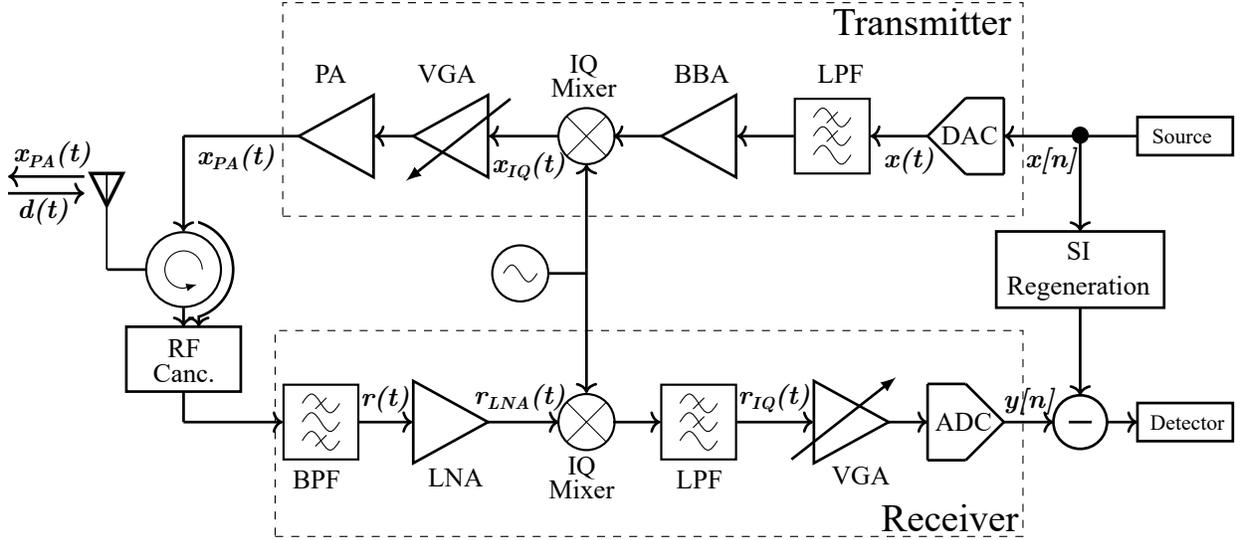}
    \caption{Detailed block diagram of a single-antenna full-duplex direct conversion transceiver}
    \label{fig: transceiver}
\end{figure*}

In this paper, we offer a comprehensive self-interference model for single-antenna full-duplex systems based on direct-conversion structure considering all the impairments including transmitter and receiver IQ imbalances, nonlinear distortions in all the transceiver components, receiver noise figure, and phase noise effect of both transmitter and receiver IQ mixers. Through extensive circuit-level waveform simulation, we show the effect of receiver RF and BB second and third-order nonlinearity on the residual self-interference signal for limited antenna and RF cancellation. To tackle this effect, we propose an appropriate self-interference estimation and digital cancellation approach considering receiver chain nonlinearities, which outperforms the existing digital cancellation techniques. To reduce the estimation and cancellation error, we employ orthogonaliztion of the design matrix using QR decomposition method.

\section{Full-duplex Transceiver Model And Self-interference characterization}\label{sec: 2}

In this section, we discuss the self-interference signal modeling for a single-antenna full-duplex transceiver system based on the direct-conversion structure presented in Fig. \ref{fig: transceiver}. In direct-conversion structure, RF and BB signal frequency translation is performed without any intermediate frequency stage unlike superheterodyne structure\cite{gu2005rf}. The direct-conversion receiver structure allows signal amplification and filtering at the BB stage, which reduces the power consumption and manufacturing cost. However, direct-conversion radios suffer from RF and BB impairments, such as IQ imbalances, nonlinear distortions, and phase noise effect \cite{schenk2008rf}. Therefore, a detailed characterization of the self-interference signal is performed in the next subsections considering the RF and BB impairments.\par

\subsection{Self-interference Model at the Transmitter}
At the transmit chain, the impairments are mainly introduced by the RF front-end components, specifically IQ mixers and power amplifiers. Both baseband and bandpass modeling approach are taken into account for the following SI characterization.\par

After the digital-to-analog converters (DACs), the I and Q components of the converted baseband signal $x(t)$ are passed through the low-pass filters for further suppression of aliasing products. The I and Q signals are then fed into the IQ mixer for upconversion to the carrier frequency. As stated in \cite{korpi2014widely}, IQ imbalances in practical mixers add a mirror image of the original signal with certain image attenuation. The IQ mixers also induce phase noise to the signal. Let $\gamma_{\scriptscriptstyle Tx}, \lambda_{\scriptscriptstyle Tx}$ be the complex gain of the linear and image signal components, respectively. Thus, the upconverted RF signal is given by
\begin{equation}\label{eq1} 
    \begin{split}
    x_{\scriptscriptstyle IQ}(t)&\doteq 2\Re\Big\{\big(\gamma_{\scriptscriptstyle Tx} ~x(t)+\lambda_{\scriptscriptstyle Tx} ~\bar{x}(t)\big)e^{j(\omega_ct+\theta_{\scriptscriptstyle Tx})}\Big\}\\
    &= z(t) e^{j(\omega_c t+\theta_{\scriptscriptstyle Tx})} + \bar{z}(t)e^{-j(\omega_c t+\theta_{\scriptscriptstyle Tx})},
    \end{split}
\end{equation}
where $z(t)=\big(\gamma_{\scriptscriptstyle Tx} x(t)+\lambda_{\scriptscriptstyle Tx} \bar{x}(t)\big)$, $\omega_c$ is the angular center frequency of the RF signal and $\theta_{\scriptscriptstyle Tx}(t)$ is the random phase noise process. Here, $(\bar{\cdot})$ denotes the complex conjugate.\par
Before transmission, the upconverted signal goes through variable amplification stages (variable gain amplifier (VGA), PA driver) to meet the transmit power specifications. Then, for final amplification, the signal is fed into the power amplifier operating in its nonlinear region.
Only the odd-order nonlinear terms are considered for PA, as the even-power harmonics lie out of band and will be cutoff by the RF low pass filter (LPF) of the receiver\cite{korpi2014widely}\cite{bharadia2013full}.
Considering the VGA gain as $\beta_{\scriptscriptstyle VGA}$, the PA nonlinear response is modeled based on the Hammerstein nonlinearity, given as
\begin{equation}\label{eq2}
    \begin{split} 
    x_{\scriptscriptstyle PA}(t)& \doteq \Big (\sum_{p\in\{1,3\}} \beta_{\scriptscriptstyle PA,p} \big(\beta_{\scriptscriptstyle VGA}~ x_{\scriptscriptstyle IQ}(t)\big)^p\Big) \ast f(t),\\
    \end{split} 
\end{equation}
where $p$ is the order of the nonlinearity, $f(t)$ is the memory polynomial, and $(\ast)$ denotes the linear convolution. Here, $\beta_{\scriptscriptstyle PA,1}$ is the linear gain and $\beta_{\scriptscriptstyle PA,3}$ is the gain of third-order nonlinear distortions. We consider only up to the third-order distortion, as that is in practice always the strongest nonlinearity at PA output\cite{korpi2014widely}. A general definition of the nonlinear distortion gains is given as
\begin{equation}\label{eq3}
    \beta_{C,n} \doteq \frac{\beta_{C,1}}{(IIPn)^{n-1}},
\end{equation}
where $\beta_{C,1}$ denotes the linear gain and $n$ is the nonlinearity order of any component $C=\{PA, LNA, BB\}$. These nonlinear distortions are modeled using IIP2 and IIP3 figures (second and third order input-referred intercept points) of the PA\cite{gu2005rf}.\par 

In single-antenna system, the PA output is routed to the antenna using a circulator. In addition to routing signals between the transceiver and antenna, the circulator should also provide a very high isolation between the transmitter (Tx) and receiver (Rx). Unfortunately, practical circulators provide only about $15-20$ dB isolation contributing to a direct path between Tx and Rx as circulator leakage\cite{bharadia2013full}. There is another path of SI due to the antenna reflection\cite{khaledian2018inherent}. Because of the mismatch between transmitter line impedance and antenna impedance, a portion of the transmitted signal reflects back to the receiver front end. While these two signals are considered to be the main coupling components of the SI, there are also weaker multipath components due to the reflections from surrounding environment. At the receiver input, the SI signal subjects to the RF cancellation. Letting $\alpha_{RF}(t)$, $h_{ch}(t)$ be the impulse response of the RF cancellation and the multipath channel response of the self-interference signal, respectively, the signal at the receiver input is given by
\begin{equation}\label{eq4}
    \begin{split}
    r(t)& \doteq \alpha_{RF}(t)\ast h_{ch}(t) \ast x_{\scriptscriptstyle PA}(t) + d(t)+ \eta_{th}(t)\\
    &\stackrel{\text{(a)}}{=}\sum_{p\in\{1,3\}} h_p (t) \ast \big(x_{\scriptscriptstyle IQ}(t)\big)^p +d(t) +\eta_{th}(t),
    \end{split}
\end{equation}
where $ h_p (t)= \beta_{\scriptscriptstyle PA,p} {\beta^{^p}_{\scriptscriptstyle VGA}}\big(\alpha_{\scriptscriptstyle RF}(t)\ast f(t) \ast h_{ch}(t)\big) $ and $d(t),\eta_{th}$ are the desired signal and the thermal noise of the receiver, respectively. Here, $(a)$ is defined using \eqref{eq2}.

\subsection{Self-interference Model at the Receiver}
As a direct-conversion receiver suffers from Low Noise Amplifier (LNA) nonlinearities, Mixer IQ imbalance, and BB nonlinearity, these impairments will have a significant effect on the self-interference signal, especially in higher transmit power case. Therefore, a detailed receiver chain SI modeling is performed considering all the receiver impairments.\par

At the first stage of the receiver, the signal is amplified by a nonlinear LNA. We model the LNA using the same Hammerstein model used in \eqref{eq3}. The LNA output is given by
\begin{equation}\label{eq5}
    \begin{split}
    r_{\scriptscriptstyle LNA}(t)\!\!& \doteq \sum_{q\in\{1,3\}} \beta_{\scriptscriptstyle LNA,q}~ \left(r(t)\right)^q+ \eta_{\scriptscriptstyle LNA} (t)\\
    &\stackrel{\text{(b)}}{=}\!\!\!\sum_{q\in\{1,3\}}\!\!\! \beta_{\scriptscriptstyle LNA,q}\Big(\!\!\! \sum_{p\in\{1,3\}}\!\!\! h_p (t)\ast \big(x_{\scriptscriptstyle IQ}(t)\big)^p\!+\!d(t)\!+\!\eta_h(t)\Big)^q\\&\qquad\qquad\qquad\qquad\qquad\qquad\qquad\qquad+\eta_{\scriptscriptstyle LNA} (t)\\
    & \approx\!\!\! \sum_{q\in\{1,3\}} \beta_{\scriptscriptstyle LNA,q}\Big( \sum_{p\in\{1,3\}} h_p (t) \ast \big(x_{\scriptscriptstyle IQ}(t)\big)^p\Big)^q + c(t),
    \end{split}
\end{equation}
where $\eta_{\scriptscriptstyle LNA}(t)$ is the noise of the LNA and $q$ is the order of the nonlinearity. Here, $(b)$ is defined using \eqref{eq4}. For brevity, we only consider the linear operation of the receiver components for the desired signal and noise. Therefore, $c(t)=\beta_{\scriptscriptstyle LNA,1}\big( d(t)+\eta_{th}(t)\big)+\eta_{\scriptscriptstyle LNA} (t)$. Here, $\beta_{\scriptscriptstyle LNA,1}$ and $\beta_{\scriptscriptstyle LNA,q}$ are the linear gain and respective $q$th order nonlinear distortion gain of the LNA. Only odd order distortions are considered, as the even order RF LNA nonlinearities produce frequency components that are far away from $\omega_c$ and will be filtered out by low pass filter (LPF)\cite{gu2005rf},\cite{schenk2008rf}.\par
After the LNA, receiver IQ mixer downconverts the RF signal into baseband frequency and LPFs are used to filter out the high frequency terms. Practical IQ mixers induce random phase noise $\theta_{\scriptscriptstyle Rx}(t)$, which is  uncorrelated to $\theta_{\scriptscriptstyle Tx}(t)$ in case of independent oscillators\cite{sahai2013impact}. However, in single-antenna full-duplex system, same oscillator is shared by the transmitter and the receiver resulting in a common phase noise process, $\theta(t)$. Thus, $\theta_{\scriptscriptstyle Tx}(t)=\theta_{\scriptscriptstyle Rx}(t)=\theta(t)$. It is shown in previous literature that sharing oscillators suppresses the phase noise below the noise floor even in the case of transmission delay \cite{balatsoukas2015baseband},\cite{sahai2013impact}. So, the downconverted signal is written as
\begin{equation}\label{eq6}
    \begin{split}
        r_{\scriptscriptstyle IQ}(t)\doteq&\text{LPF}\big\{r_{\scriptscriptstyle LNA}(t)e^{-j(\omega_c t+ \theta_{Rx}(t))}\big\}\\
        =&\text{LPF}\Big\{\!\bigg(\!\!\sum_{q\in\{1,3\}}\!\!\beta_{\scriptscriptstyle LNA,q}\Big(\!\!\sum_{p\in\{1,3\}}\!\! h_p (t)\! \ast\! \big(x_{\scriptscriptstyle IQ}(t)\big)^p\Big)^q\!\!+c(t)\!\!\bigg)\\&\qquad\qquad\qquad\qquad\quad\qquad\qquad\quad\;\;\;
        e^{-j(\omega_c t+ \theta_{Rx}(t))}\Big\}\\
        =&\text{LPF}\Big\{\!\sum_{q\in\{1,3\}}\!\beta_{\scriptscriptstyle LNA,q}\Big(\!\!\sum_{ p\in\{1,3\}}\!\! h_p (t)\! \ast\! \big(z(t)e^{j(\omega_c t+ \theta_{Rx}(t))}\\& \quad+\bar{z}(t)e^{-j(\omega_c t+ \theta_{Rx}(t))}\big)^p\Big)^q
        e^{-j(\omega_c t+ \theta_{Rx}(t))}\Big\}+c(t)\\
        =&\!\!\!\!\sum_{q\in\{1,3\}}\!\!\!\beta_{\scriptscriptstyle LNA,q}\!\left(\!\!\!\!\!\begin{array}{c}
q \\
\frac{q-1}{2}
\end{array}\!\!\!\!\!\right)\!\! \big(h_1(t)\!\ast\! z(t)\!+ \!h_3(t)\! \ast\!  3 z^2(t)\bar{z}(t)\big)^{\frac{q+1}{2}}\\&\qquad\quad \big(h_1(t)\ast \bar{z}(t)+ h_3(t)\ast 3 z(t){\bar{z}(t)}^2\big)^{\frac{q-1}{2}}+c(t) \\
        =& \sum_{q\in\{1,3\}}\beta_{\scriptscriptstyle LNA,q}\left(\!\!\!\!\!\begin{array}{c}
q \\
\frac{q-1}{2}
\end{array}\!\!\!\!\!\right) \Big(\sum_{p\in\{1,3\}}h_p(t)\ast u_p(t)\Big)^{{\frac{q+1}{2}}}\\&\qquad\qquad\qquad \Big(\sum_{p\in\{1,3\}}h_p(t)\ast \bar{u}_p(t)\Big)^{{\frac{q-1}{2}}}+c(t)\\
    =& s_{IQ}(t) + c(t),
    \end{split}
\end{equation}
where
\begin{equation}\label{eq7}
    \begin{split}
        s_{IQ}(t)\!=& \sum_{q\in\{1,3\}}\beta_{\scriptscriptstyle LNA,q}
        \binom{q}{\frac{q-1}{2}}
        \Big(\sum_{p\in\{1,3\}}h_p(t)\ast u_p(t)\Big)^{{\frac{q+1}{2}}}\\&\quad\qquad\qquad\qquad\qquad\quad\Big(\sum_{p\in\{1,3\}}h_p(t)\ast \bar{u}_p(t)\Big)^{{\frac{q-1}{2}}}\\
        u_1(t)\!=& \gamma_{\scriptscriptstyle Tx} ~x(t)+\lambda_{\scriptscriptstyle Tx} ~\bar{x}(t)\\
        u_3(t)\!=&3\Big(\!\! \gamma_{\scriptscriptstyle Tx}^2 \bar{\lambda}_{\scriptscriptstyle Tx} ~x^3(t)\! +\! (\gamma_{\scriptscriptstyle Tx}^2 \bar{\gamma}_{\scriptscriptstyle Tx}\! +\! 2 \gamma_{\scriptscriptstyle Tx} \lambda_{\scriptscriptstyle Tx} \bar{\lambda}_{\scriptscriptstyle Tx}) x^2(t)\bar{x}(t)\\&\!\! + \!(2 \gamma_{\scriptscriptstyle Tx} \bar{\gamma}_{\scriptscriptstyle Tx}\lambda_{\scriptscriptstyle Tx}  \!+\! \lambda_{\scriptscriptstyle Tx}^2 \bar{\lambda}_{\scriptscriptstyle Tx})~ x(t)\bar{x}^2(t)\! + \!\lambda_{\scriptscriptstyle Tx}^2 \bar{\gamma}_{T\scriptscriptstyle x} 
        \bar{x}^3(t)\Big).\\
    \end{split}
\end{equation}
 The reader is referred to the Appendix for necessary proofs of \eqref{eq6}.  
Receiver IQ mixer also induces the image component because of the IQ imbalance. Now, considering the effect of IQ imbalance, the IQ output signal is written as
\begin{equation}\label{eq8}
    \begin{split}
    r_{\scriptscriptstyle IQIm}(t)\doteq\gamma_{Rx}~r_{\scriptscriptstyle IQ}(t)+\lambda_{Rx}~ \bar{r}_{\scriptscriptstyle IQ}(t)+\eta_{\scriptscriptstyle IQ}(t),
    \end{split}
\end{equation}
where $\gamma_{\scriptscriptstyle Rx},\lambda_{\scriptscriptstyle Rx}$ are  the linear and image component gain, and $\eta_{\scriptscriptstyle IQ}(t)$ is the noise of the IQ mixer.\par
\begin{table*}[b]
    % \begin{tabular}{c}
    \noindent\makebox[\linewidth]{\rule{\textwidth}{0.4pt}}
    % \hline
    \normalsize
    \begin{equation}\label{eq12}\tag{12}
        \begin{split}
            \boldsymbol \Psi &=\boldsymbol{\begin{bmatrix} 1 & \Psi_{x} & \Psi_{\bar{x}} &\Psi_{x,x}& \Psi_{x,\bar{x}}& \Psi_{\bar{x},\bar{x}} & \Psi_{x,x,x}& \Psi_{x,x,\bar{x}}& \Psi_{x,\bar{x},\bar{x}}& \Psi_{\bar{x},\bar{x},\bar{x}} & \Psi_{x,x,x^2\bar{x}} & \Psi_{x,\bar{x},x^2\bar{x}} & \Psi_{x,x,x\bar{x}^2}\end{bmatrix}},
        \end{split}
    \end{equation}
    where $\boldsymbol{1}$ denotes an $N\times1$ column vector with all ones. Letting $\boldsymbol{a,b,c \in \{x,\bar{x},x^2\bar{x},x\bar{x}^2\}}$, each of the horizontally concatenated matrices of $\boldsymbol{\Psi}$ can be formulated as 
    \begin{equation}\label{eq13}\tag{13}
        \begin{split}
            \boldsymbol{\Psi_a}&=\begin{bmatrix}\Phi_{\boldsymbol a}(n) & \Phi_{\boldsymbol a}(n-1)&\dots & \Phi_{\boldsymbol a}(n-L+1)\end{bmatrix}
            \\
            \boldsymbol{\Psi_{a,b}}&=\begin{bmatrix}\Phi_{\boldsymbol b}(n)\odot\boldsymbol{\Psi_{a}} & \Phi_{\boldsymbol b}(n-1)\odot\boldsymbol{\Psi_{a}}&\dots & \Phi_{\boldsymbol b}(n-L+1)\odot\boldsymbol{\Psi_{a}}\end{bmatrix}\\
            \boldsymbol{\Psi_{a,b,c}}&=\begin{bmatrix}\Phi_{\boldsymbol c}(n)\odot\boldsymbol{\Psi_{a,b}} & \Phi_{\boldsymbol c}(n-1)\odot\boldsymbol{\Psi_{a,b}}&\dots & \Phi_{\boldsymbol c}(n-L+1)\odot\boldsymbol{\Psi_{a,b}}\end{bmatrix},
        \end{split}
    \end{equation}
    where $\odot$ denotes element wise multiplication of the vector with all the columns of the matrix. Here, the basis vectors are defined as
    \begin{equation}\label{eq14}\tag{14}
        \begin{split}
            \Phi_{\boldsymbol {x}}(n)&=\begin{bmatrix}x[n] & x[n+1] & \dots & x[n+N-1]\end{bmatrix}^T, \qquad\qquad\qquad\qquad\qquad\quad\qquad\qquad\;\,
            \Phi_{\boldsymbol {\bar{x}}}(n)=\bar{\Phi}_{\boldsymbol {x}}(n),\\
            \Phi_{\boldsymbol {x^2}\boldsymbol {\bar{x}}}(n)&=\begin{bmatrix}x^2[n]\bar{x}[n] & x^2[n+1] \bar{x}[n+1] &\dots & x^2[n+N-1]\bar{x}[n+N-1]\end{bmatrix}^T, \qquad \Phi_{\boldsymbol{x}\boldsymbol{\bar{x}^2}}(n)=\bar{\Phi}_{\boldsymbol {x^2}\boldsymbol {\bar{x}}}(n).
        \end{split}
    \end{equation}
    % \end{tabular}
\end{table*}
Baseband components shown in Fig. \ref{fig: transceiver}, specifically amplifiers and analog-to-digital converters (ADCs) also introduce some nonlinearities and DC offset to the BB signal. For simplification, up to second order nonlinearity is considered. In our model, we use the complex representation of the baseband signal since it helps analytically by revealing how the distortion components are spectrally distributed in relation to the original signal. Considering all the impairments, the complex baseband signal is written as 
\begin{equation}\label{eq9}\tag{9}
    r_{\scriptscriptstyle BB}(t)\doteq\sum_{m=0}^2 \beta_{\scriptscriptstyle BB,r}~ r_{\scriptscriptstyle IQIm}^m (t) + \eta_{\scriptscriptstyle BB}(t),
\end{equation}
where $m$ is the order of the nonlinearity, $\beta_{\scriptscriptstyle BB,r}$ and $\eta_{\scriptscriptstyle BB}(t)$ is the respective $m$th order  baseband gain and the BB components noise, respectively. The first term of \eqref{eq9} represents the DC offset introduced by the baseband components.
At the end of the receiver chain, the ADC converts the analog baseband signal to digital domain. Considering the quantization noise, $n_{\scriptscriptstyle quant}[n]$ induced by the ADC, the digital signal at the ADC output is expressed as
\begin{equation}\label{eq10}\tag{10}
    \begin{split}
    y[n]\doteq& \;r_{\scriptscriptstyle BB}[n] + \eta_{\scriptscriptstyle Quant} [n]\\ =& \sum_{m=0}^2 \beta_{\scriptscriptstyle BB,r}~ \Big(\gamma_{\scriptscriptstyle Rx}~ s_{\scriptscriptstyle IQ}[n]+\lambda_{\scriptscriptstyle Rx}~ \bar{s}_{\scriptscriptstyle IQ}[n]\Big)^m\\&\qquad\qquad+\beta_{\scriptscriptstyle BB,1}~\gamma_{\scriptscriptstyle Rx}~\beta_{\scriptscriptstyle LNA,1}~ d[n]+ \eta_T[n] .
    \end{split}
\end{equation}
Here, $\eta_T[n]$ represents the total noise of the receiver including thermal noise, receiver noise figure and quantization noise.
\section{Proposed Digital Estimation and Cancellation of Self-interference Signal}

In this section, we propose an appropriate digital cancellation approach with improved reference signal design for estimation based on the signal at the ADC output. To ensure the estimation accuracy, orthogonalization procedure is employed in estimation technique.\par 

Based on \eqref{eq10} and \eqref{eq7}, $y[n]$ can be expressed as the combination of the linear SI signal $x[n]$, conjugate SI signal $\bar{x}[n]$, and their higher order terms along with desired signal, and receiver noise floor. Although a detailed expansion of \eqref{eq10} results in hundreds of residual SI terms, most of the higher order terms are very weak. Therefore, they have insignificant effect on the total self-interference power, thus can be ignored without compromising self-interference cancellation accuracy. Based on the specification in Table \ref{tab:2}, we keep the 13 stronger SI components that include combination of $x[n]$, $\bar{x}[n]$, and their higher order terms and use those for cancellation of the self-interference signal.
Hence, using vector matrix notation, for $N$ observed training samples, the ADC output signal is written as 
\begin{equation}\label{eq11}\tag{11}
    \begin{split}
        \boldsymbol{Y}= \boldsymbol{\Psi}\boldsymbol{w} + \boldsymbol{d} + \boldsymbol{\eta},
    \end{split}
\end{equation}
where $\boldsymbol Y=\begin{bmatrix} y[n] & y[n+1]& \dots & y[n+N-1]\end{bmatrix}^T$, $\boldsymbol d= \beta_{\scriptscriptstyle BB,1}\gamma_{\scriptscriptstyle Rx}\beta_{\scriptscriptstyle LNA,1} \begin{bmatrix}d[n]& d[n+1]& \dots & d[n+N-1]\end{bmatrix}^T$, and $\boldsymbol \eta=\begin{bmatrix} \eta_T[n] & \eta_T[n+1]& \dots & \eta_T[n+N-1]\end{bmatrix}^T$.
% Here, the desired signal $\boldsymbol d$ and noise $\boldsymbol{\eta}$ has the same structure as $\boldsymbol Y$.
The definition of the design matrix $\boldsymbol{\Psi}$ with size $N\times(2L+3L^2+7L^3+1)$ are given in \eqref{eq12}, where $L$ is the length of the total impulse response including multipath channel, analog cancellation, and PA memory polynomial. Here, the horizontally concatenated basis matrices represent the 13 stronger SI terms formulated using \eqref{eq13} and \eqref{eq14}. For instance, the matrix $\boldsymbol{\Psi_{x,\bar{x},x^2\bar{x}}}\in \mathbb{C}^{N\times L^3}$ represent the SI term:
$9\beta_{\scriptscriptstyle BB,1}\gamma_{\scriptscriptstyle Tx}\beta_{\scriptscriptstyle LNA,3}\gamma_{\scriptscriptstyle Rx}\bar{\gamma}_{\scriptscriptstyle Tx}(\gamma_{\scriptscriptstyle Tx}^2 \bar{\gamma}_{\scriptscriptstyle Tx} + 2 \gamma_{\scriptscriptstyle Tx} \lambda_{\scriptscriptstyle Tx} \bar{\lambda}_{\scriptscriptstyle Tx})\big(h_1[n]\ast x[n]\big)\big( h_1[n] \ast \bar{x}[n]\big)\big(h_3[n]\ast x^2[n]\bar{x}[n]\big)$.\par 

Here, $\boldsymbol{w}$ is the $(2L\!+\!3L^2\!+\!7L^3\!+\!1)\!\times\!1$ parameter vector.
Our goal is to estimate the parameter $\boldsymbol{w}$, and then use it to reconstruct and cancel the self-interference signal at the detector input. Therefore, the error vector is defined as 
\begin{equation}\label{eq15}\tag{15}
    \begin{split}
        \mathbf{e}=\mathbf{Y}-\boldsymbol{\Psi\hat{w}},
    \end{split}
\end{equation}
where $\boldsymbol{\hat{w}}$ is the least square estimate, which can be obtained by solving for the $\boldsymbol{w}$ that minimizes the power of the error vector as
\begin{equation}\label{eq16}\tag{16}
    \boldsymbol{\hat{w}}=\argmin_{\boldsymbol{w}} \norm{\mathbf{e}}^2=\argmin_{\boldsymbol{w}} \norm{\mathbf{\mathbf{Y}-\boldsymbol{\Psi w}}}^2.
\end{equation}\par
The above ordinary least square problem has closed form solution with the usual assumption that the design matrix $\boldsymbol{\Psi}$ has independent columns. However, the columns of the matrix $\boldsymbol{\Psi}$ are higher order polynomials of linear SI component $x[n]$, conjugate SI component $\bar{x}[n]$, and their interaction products. Therefore, the columns of the matrix  $\boldsymbol{\Psi}$ are correlated, thus there exists high level of multicollinearity. To tackle this problem, we propose to orthogonalize the design matrix $\boldsymbol{\Psi}$ using QR decomposition. We use traditional Gram-Schmidt orthogonalization procedure to find a QR decomposition of $\boldsymbol{\Psi}$ such that $\boldsymbol{\Psi}=\mathbf{QR}$, where $\mathbf{Q}$ has orthogonal columns and same rank as  $\boldsymbol{\Psi}$. Here, $\mathbf{R}$ is an upper triangular matrix with 1 on the diagonal. Using the QR decomposition, \eqref{eq16} is written as
\begin{equation}\label{eq17}\tag{17}
    \begin{split}
        \boldsymbol{\hat{w}}=\argmin_{\boldsymbol{w}} \norm{\mathbf{\mathbf{Y}-\mathbf{QR}\boldsymbol{w}}}^2.
    \end{split}
\end{equation}
Letting $\boldsymbol{\mu}=\mathbf{R}\boldsymbol{w}$, we can reformulate the above minimization problem and find the least square solution as
\begin{equation}\label{eq18}\tag{18}
    \begin{split}
    \boldsymbol{\hat{\mu}}=&\argmin_{\boldsymbol{\mu}} \norm{\mathbf{\mathbf{Y}-\mathbf{Q}\boldsymbol{\mu}}}^2\\
    =&\mathbf{(Q^HQ)^{-1}Q^H Y},
    \end{split}
\end{equation}
where $(\cdot)^H$ denotes the Hermitian transpose of the matrix. Using back substitution, we derive $\boldsymbol{\hat{w}}$ from $\boldsymbol{\hat{\mu}}$.\par 

It is to be noted that authors in \cite{korpi2014widely} proposed a widely linear canceller considering only the effect of linear SI $x[n]$ and conjugate SI $\bar{x}[n]$, a nonlinear approach considering third order nonlinear SI term $x^2[n]\bar{x}[n]$ is provided in \cite{bharadia2013full}, and a joint cancellation technique
% $x^3,x^2\bar{x}$, and $x\bar{x}^2$ 
cascading PA nonlinearity with transmitter IQ imbalance
is proposed in \cite{anttila2014modeling} for MIMO cases, which is converted here into single antenna system. However, receiver nonlinearities and its effect are ignored in these previous models. Our proposed estimation approach takes these components into account for the formulation of the design matrix and provides required SI cancellation at the detector input.

\section{Perfromance Simulation and Analysis}

In this section, we perform a waveform simulation to evaluate the performance of the proposed digital cancellation strategy and compare with the previous methods. 
\begin{table}[tbp]
    \caption{Waveform and System Level Parameters}
    \label{tab:1}
    \centering
    \scriptsize
    \begin{tabular}{|c||c|}
         \hline
         \textbf{Parameter} & \textbf{Value} \\
         \hline
         Bandwidth & 20 MHz \\
         Carrier Frequency & 2.4 GHz\\
         Constellation & 16-QAM\\
         Number of Subcarriers &  64\\
         Cyclic Prefix Length & 16\\
         Sample length & 15.625 ns\\
         Symbol Length & 4 $\mu$s\\
         \hline
    \end{tabular}
    \,
    \begin{tabular}{|c ||c|}
         \hline
         \textbf{Parameter} & \textbf{Value} \\
         \hline
         SNR Requirement & 10 dB\\
         Thermal Noise Floor & $-$101.0 dBm\\
         RX Noise Figure & 4.1 dB\\
         Sensitivity Level & $-$86 dBm\\
         Transmit Power & $-$5$:$25 dBm\\
         ADC Bits & 12\\
         PAPR & 10 dB\\
         \hline
    \end{tabular}
\end{table}
% \vspace{1mm}
\begin{table}[tbp]
    \caption{Full-duplex Transceiver Component Parameters}
    \label{tab:2}
    \centering
    \scriptsize
    \begin{tabular}{|c||c|}
         \hline
         \textbf{Parameter} & \textbf{Value} \\
         \hline
         PA Gain & 27 dB \\
         PA IIP3 & 13 dBm\\
         VGA (Tx) Gain & 0$:$33 dB\\
         IRR (Tx) & 30 dB\\
         \hline
    \end{tabular}
    \,
    \begin{tabular}{|c||c|}
         \hline
         \textbf{Parameter} & \textbf{Value} \\
         \hline
         LNA Gain & 20 dB \\
         LNA IIP3 & $-$3 dBm\\
         BB IIP2 & 50 dBm\\
         IRR (Rx) & 30 dB\\
         \hline
    \end{tabular}
\end{table}
\normalsize
\subsection{ Simulation Parameters}
The transceiver architecture provided in Fig. \ref{fig: transceiver} is followed explicitly to perform the waveform simulation. The simulator is implemented using MATLAB baseband equivalent model of the components, which includes the impact of all the transceiver impairments. The waveform simulation is based on a 20 MHz OFDM-based system with 64 subcarriers per OFDM symbols as in IEEE802.11 systems. The additional parameters of the waveforms are shown in Table \ref{tab:1}. The system level component parameters of the full-duplex transceiver are provided in Table \ref{tab:2}, which are taken from the datasheet of AD9361 transceiver made by Analog Devices\cite{ad9361}. We perform our simulation based on single-antenna full-duplex system with a circulator providing a stronger leakage signal to the receiver as the direct SI component and several multipath components\cite{debaillie2014analog}. Therefore, the SI channel is simulated as a Rician channel with a K-factor of 30 dB and 4 non-line-of-sight (nLOS) components each delayed by one sample\cite{duarte2012experiment}. To emulate IEEE802.11 system, we use WLAN packet formatting where each packet consists of 120 OFDM symbols. Around 8$\%$ of the total packet size are considered as training symbols, which are used to estimate and cancel the self-interference. We use 1000 Monte Carlo simulation runs to calculate the average cancellation performance. Here, we consider 15dB circulator isolation with variable RF cancellation values to show its impact on the digital cancellation.\par

\begin{figure}[!tbp]
\centering
\includegraphics[width=0.46\textwidth]{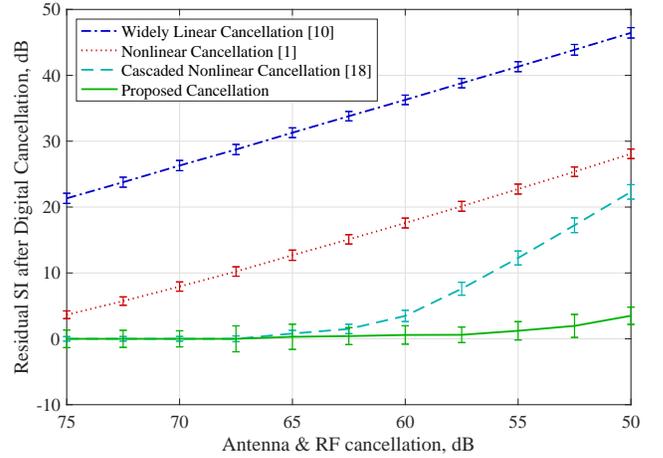}
\caption{Residual self-interference power after digital cancellation with respect to different RF cancellation value and 25 dBm transmit Power}
\label{fig: signal}
\end{figure}
\subsection{SI Cancellation Capability and SINR}
First, for variable antenna and RF cancellation, we compare the performance of the proposed digital cancellation technique with previous methods. 
In Fig. \ref{fig: signal}, we represent the residual self-interference signal strength of several digital cancellation methods from the measured noise floor with respect to 50-75 dB antenna and RF cancellation for a transmit power of 25 dBm. We also include the error bars at one standard deviation from the mean of the cancellation value. We observe that widely linear cancellation\cite{korpi2014widely} scheme has a residual of 20-45 dB. Although considering the nonlinear cancellation\cite{bharadia2013full} along with linear and widely linear methods increases the performance drastically, it has a residual of around 30 dB at an antenna and RF cancellation of 50 dB. For higher antenna and RF suppression, cascaded cancellation\cite{anttila2014modeling} performance is comparable with our proposed method. However, below 65 dB cancellation, it has a sharp growth of the residual leading almost 22 dB for 50 dB antenna and RF cancellation. In comparison with these methods, our proposed digital cancellation technique provides better performance contributing only around 3 dB residual at a very low RF cancellation. This performance boost is because of the inclusion of receiver chain second and third-order nonlinearities in our proposed method, whereas all the previous methods took only the transmitter PA nonlinearities into account ignoring the receiver chain nonlinear distortions. However, that assumption provides comparable performance in case of high antenna and RF cancellation or linear receiver operation. As we observed, for practical  transceivers such as AD9361, the receiver performance is not linear. Therefore, limited antenna and RF cancellation, typically below 65 dB, results in the higher self-interference signal power at the receiver input, which contributes to a stronger nonlinear effect of the receiver chain components. In this scenario, the proposed digital cancellation strategy outperforms the previous techniques.\par

\begin{figure}[!tp]
\centering
\includegraphics[width=0.46\textwidth]{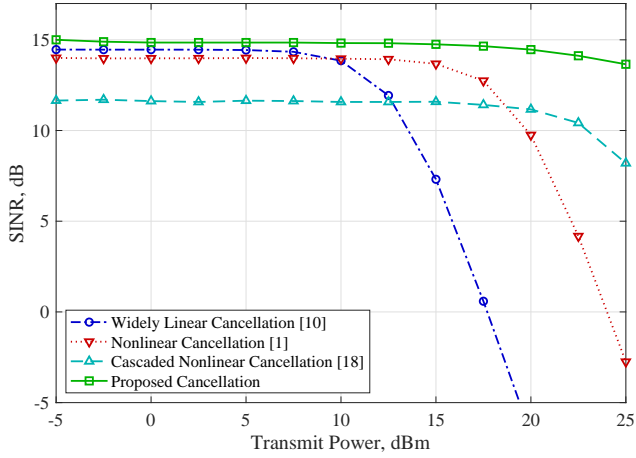}
\caption{The SINRs for different digital cancellation method with respect to the overall transmit power with antenna and RF cancellation of 60 dB}
\label{fig: signal_tx_power}
\end{figure}
In Fig. \ref{fig: signal_tx_power}, we show the performance of the digital cancellation methods with respect to the transmit power varying from $-$5 dBm to 25 dBm while considering 60 dB antenna and RF cancellation. Here, signal-to-interference-plus-noise-ratio (SINR) is used as the figure-of-merit of the cancellation strategies. During calculation of the SINR, the desired signal is naturally present with an SNR of 15 dB, which is the maximum achievable SINR. From Fig. \ref{fig: signal_tx_power}, we observed that our proposed method outperforms the previous techniques for all the transmit power cases. Nonlinear\cite{bharadia2013full} and widely linear\cite{korpi2014widely} methods sufficiently suppress the SI up to the transmit power of 15 and 10 dBm, respectively. Although, the cascaded cancellation\cite{anttila2014modeling} method performs better than the previous methods for higher transmit power cases, it suffers from overfitting problem resulting in higher estimation error in low transmit power cases. In our proposed model, we tackle this problem through orthogonalization of the design matrix, which reduces the estimation error. Thus, our proposed digital cancellation approach provides reliable and sufficient suppression of self-interference signal in a practical nonlinear receiver for a varying antenna and RF cancellation as low as 50 dB.\par 

\section{Conclusion}
The performance of self-interference cancellation in single-antenna full-duplex systems is limited by the presence of hardware impairments in the transceiver chain. In this paper, we provided a detailed modeling of the self-interference signal considering transmitter PA nonlinearities, receiver chain second and third-order nonlinearities, DC offset, IQ imbalances, phase noise, and receiver noise figure. To validate the self-interference model, we proposed a digital cancellation approach considering receiver chain second and third-order nonlinearities and implemented a practical waveform simulator. Our comprehensive simulation showed that, in practical receiver, these nonlinearities have significant effect on the self-interference signal for an analog and RF cancellation of 65 dB or below. In this scenario, our proposed digital cancellation technique outperforms the existing cancellation methods by achieving up to 20 dB more self-interference cancellation. For future work, we intend to extend the self-interference modeling and cancellation for MIMO system considering all the practical transceiver impairments and provide an efficient digital self-interference cancellation strategy.
\section*{Appendix}
Necessary proofs for \eqref{eq9} are provided here.\\ Let $\omega_c t+ \theta_{Rx}(t)=\omega_c t+ \theta_{Tx}(t)=\Delta(t)$ as $\theta_{Rx}(t)=\theta_{Tx}(t)$.\\
Now,
\begin{equation}\label{eq19}\tag{19}
    \begin{split}
    % r_{\scriptscriptstyle IQ}(t)&
   &\text{LPF}\Big\{\bigg(\sum_{q=1,3}\!\!\beta_{\scriptscriptstyle LNA,q}\Big( \sum_{p=1,3}\!\! h_p (t)\! \ast\! \big(x_{\scriptscriptstyle IQ}(t)\big)^p\Big)^q\!\!+c(t)\!\!\bigg) e^{-j\Delta(t)}\Big\}\\
    &=\text{LPF}\Big\{\!\!\!\sum_{q=1,3}\!\!\!{\beta_{\scriptscriptstyle LNA,q}} \Big(\big(h_1(t)\ast z(t)\!+\! h_3(t)\ast 3 z^2(t)\bar{z}(t)\big)e^{j\Delta(t)}\\&\quad\;+\big(h_1(t)\ast \bar{z}(t)+ h_3(t)\ast 3 z(t){\bar{z}(t)}^2\big)e^{-j\Delta(t)}\Big)^q e^{-j\Delta(t)}\Big\}\\
    &=\! \text{LPF}\Big\{\!\!\!\sum_{q=1,3}\!\!\!{\beta_{\scriptscriptstyle LNA,q}}\!\!\sum_{k=0}^q \!\!\!\left(\!\!\!\!\begin{array}{c}
q \\
k
\end{array}\!\!\!\!\right)\!\!\big(h_1(t)\!\ast\!z(t)\!+\! h_3(t)\!\ast\!3 z^2(t)\bar{z}(t)\big)^{q-k} \\&\qquad\quad\big(h_1(t)\ast \bar{z}(t)+ h_3(t)\ast 3 z(t){\bar{z}(t)}^2\big)^{k}
    e^{j\Delta(t)(q-2k-1)}\Big\}\\
    &\stackrel{\text{(c)}}{=}\!\!\!\sum_{q=1,3}\beta_{\scriptscriptstyle LNA,q}\binom{q}{\frac{q-1}{2}} \big(h_1(t)\ast z(t)+ h_3(t)\ast 3 z^2(t)\bar{z}(t)\big)^{\frac{q+1}{2}}\\&\qquad\qquad\qquad\qquad \big(h_1(t)\ast \bar{z}(t)+ h_3(t)\ast 3 z(t){\bar{z}(t)}^2\big)^{\frac{q-1}{2}}.
    \end{split}
\end{equation}
Here, $(c)$ is achieved if
\begin{equation}\label{eq20}\tag{20}
    \begin{split}
        &q-2k-1=0 \Rightarrow k= \frac{q-1}{2}.
    \end{split}
\end{equation}
\section*{Acknowledgments}
This work was partially funded by the National Science Foundation CAREER award \#1620902.

\bibliographystyle{IEEEtran}
\bibliography{mybib.bib}

\end{document}